\documentclass[11pt,twoside]{article}
\usepackage{asp2014}

\aspSuppressVolSlug
\resetcounters

\begin{document}

\title{SpectraPy: a Python library for spectroscopic data reduction.}

% full name: Marco Fumana
\author{Marco~Fumana$^1$}
\affil{$^1$INAF-IASF Milano, via Alfonso Corti 12, 20133 Milano, Italia; \email{marco.fumana@inaf.it}}

% remove/add authors as you need
\paperauthor{Marco~Fumana}{marco.fumana@inaf.it}{0000-0001-6787-5950}{}{INAF-IASF}{Milano}{}{20133}{Italy}
% remove/add as you need

\begin{abstract}

SpectraPy is an Astropy affiliated package for spectroscopic data reduction. It collects algorithms and methods for data reduction of astronomical spectra obtained by through-slits spectrographs. It has been created to fill the gap in Astropy between the already existing data handling libraries and those for spectra analysis.

SpectraPy combines Astropy facilities with SAOImageDS9 features, providing a set of tools for spectra calibration and 2D extraction. It starts from raw frames, and using configuration files which describe the optical setup of the instrument, it automatically locates and extracts 2D spectra that have been wavelength calibrated and corrected by distortions.

The library is designed to be spectrograph-independent and can be used on both longslit and multi object spectrograph data. It comes with a set of ready-to-use configuration files for the LBT-LUCI and LBT-MODS spectrographs, but it can be configured for data reduction of other through-slits spectrographs.

In the future I plan to extend SpectraPy to achieve a full data reduction for both through-slit and fiber fed spectrographs.

\end{abstract}

\section{Introduction}
   While there are AstroPy affiliated packages devoted to spectral analysis, e.g. \href{https://pyspeckit.readthedocs.io/en/latest/}{PySpecKit} \citep{PySpecKit} or \href{https://linetools.readthedocs.io/en/latest/}{linetools}, there are no packages devoted to reduction of spectroscopic data.
   
   SpectraPy \citep{Fumana} is the first brick of a Python environment for spectroscopic data reduction. It provides capabilities to perform data reduction of longslit (LS) and multi object spectrograph (MOS) data. The current implementation is focused on the 2D spectra extraction: SpectraPy starts from raw dispersed frames and finishes producing 2D rectified spectra, that have been wavelength calibrated and corrected by instrument distortions.
   
   SpectraPy is a \emph{library} entirely written in Python, with the exception  of few methods written in Cython for execution time reasons. It makes an intensive use of facilities provided by \href{https://www.astropy.org/index.html}{Astropy} \citep{AstroPy1, AstroPy2}, \href{https://numpy.org/}{Numpy} and \href{https://matplotlib.org/}{matplotlib}. It also uses \href{http://ds9.si.edu/site/Home.html}{sao DS9} \citep{DS9} to display data and \href{http://hea-www.harvard.edu/RD/pyds9/}{pyds9} to interact with it.
   
   SpectraPy was born from the experience gained by developing VIPGI \citep{VIPGI}, the reduction software used to reduce more than 145,000 spectra of the major optical extragalactic spectroscopic surveys conducted with the VIMOS spectrograph: VVDS \citep{VVDS}, VIPERS \citep{VIPERS1, VIPERS2}, VUDS \citep{VUDS}, VANDELS \citep{VANDELS1, VANDELS2}, zCOSMOS \citep{zCOSMOS}. 
  
  It also takes benefits from the experience in developing SIPGI \citep{SIPGI, F03_adass}, the official Italian reduction pipeline for data acquired with \href{http://www.mpe.mpg.de/ir/lucifer/}{LBT-LUCI} and \href{http://www.astronomy.ohio-state.edu/MODS/}{LBT-MODS} spectrographs.
   
  The current releases of the software, together with an extensive manual, is available on this \href{https://gitlab.com/mcfuman/SpectraPy/}{gitlab page}\footnote{https://gitlab.com/mcfuman/SpectraPy/}.
   
\section{The mathematical models}
  In order to properly locate and extract 2D spectra, the instrument instrument distortions and grism dispersion properties must be known. SpectraPy analytically maps these distortions using a set of three \emph{mask independent} mathematical models.
  The mask independence of the models allows SpectraPy to apply these models at different observations with the same instrument configuration.

\subsection{Location of the slits: the Optical Model}\label{par:OPT}
   The first step of the library is the proper location of the slits on the frames, but on dispersed frames the images of the slits are not visible. 
   
   SpectraPy, using a lamp frame, assumes as the slit position, the position of a bright isolated emission line (\emph{the reference lambda}) close to the grism central wavelength.
    
   SpectraPy compares this reference lambda position on the dispersed frame with the nominal slit position derived from the mask description. By this comparison, it computes the Optical Model which maps the field of view (FoV) distortions. The model maps the transformation from the millimeters of the mask $(x_{mm}, y_{mm})$, to pixels on the detector $(x_{pix},y_{pix})$.

\subsection{Spectra tracing: the Curvature Model}

   The Curvature Model describes the spectra tracing, i.e. the deviations of the spectra traces with respect to the perfect straight line. 

   For each slit, SpectraPy uses one mono dimensional polynomial ($L_a$) to describe the displacement along the cross dispersion direction ($\Delta c_{pix}$) as function of the displacement ($\Delta d_{pix}$) from slit reference position located by the Optical Model (Eq. \ref{eq:local_crv}):

   \begin{equation}
       \Delta c_{pix} = L_a(\Delta d_{pix}) = \sum_{i=0}^N a_{i}(x_{pix},y_{pix}) (\Delta d_{pix})^i.
       \label{eq:local_crv}
   \end{equation}
   The coefficients ${a_i(x_{pix}, y_{pix})}$ of the \emph{local} model, $L_a$, depend on the position $(x_{pix}, y_{pix})$ of the slit on the detector. Each slit has its own set of ${a_i}$ coefficients, because each slit is in a different position on the detector. 
   
   SpectraPy uses a \emph{global model} ($G_A$) to describe the coefficients variations along the FoV. This approach makes the model mask independent: using a global model the connection between mask and local models has been removed. 
   This implies that once $G_A$ is calibrated, the library can compute the $\{a_i\}$ set everywhere in the FoV for all data acquired with the same configuration.

\subsection{Wavelength calibration: the Inverse Dispersion Solution}\label{par:IDS}

   Once the spectra have been located on the detector by the Optical Model and geometrically described by the Curvature Model, the wavelength calibration of the 2D spectra can be carried out. The Inverse Dispersion Solution is the model used to obtain the relation between wavelengths and pixel positions. For each spectrum, this model moves along the curve described by the combination of Optical Model and Curvature Model, and it associates wavelength values to spectrum pixels: $\lambda \rightarrow (x_{pix}, y_{pix})$.
   
   The mathematical description of the Inverse Dispersion Solutions is quite similar to the Curvature Model: for each slit, one mono dimensional polynomial $L_b$ locates the wavelength position. Even in this case, the set of coefficients of $L_b$ are obtained by the evaluation of the global model $G_B$.

\section{Models calibrations}
   The calibration of the three models described above is performed in 2 steps: manual calibration and automatic refinement on data.

\altsubsubsection*{Manual calibration}
   The manual calibration is occasionally performed by the users. The first time users are working on a new set of data, they must use SpetraPy to manually tune the models. SpectraPy uses DS9 to display the frames and the \href{https://pyregion.readthedocs.io/en/latest/}{pyregions} package to display the models as DS9 regions. The users have to check these models and adjust the regions  according with the frame features.
   
   Since the SpectraPy models are designed to be mask independent, this manual adjustment is done \emph{once for all} (if the instrument is stable) for a given instrument configuration (i.e. camera, grism, dichroic). If the users acquire new data with the same instrument configuration, but different masks, the users can reuse models already calibrated, i.e. no further manual adjustments is required.

\altsubsubsection*{Automatic refinements on data}
   Usually the observations of a single target are spread out over several nights and instrument distortions can slightly change on a night basis. 
   These tiny changes can be automatically recovered by SpectraPy: no manual operations are required. 
   
   Starting from the solutions computed during the manual adjustment procedure on the data of one specific night, SpectraPy is able to automatically adjust the Curvature Model and the Inverse Dispersion Solution Model.

\section{2D spectra extraction}
   The 2D spectra extraction is the final step: once the models are automatically refined on data, the library can extract the 2D spectrum from each slit.
   The user can choose the wavelength range to extract and the re-binning step size. 
   
   The 2D extraction procedure is basically an image warping process and it involves a re-sampling function. The re-sampling procedure is the process of transforming the raw spectra from one coordinate system described by the models, to another where the spectra are perfectly rectified and wavelength calibrated.

   SpectraPy computes the re-sampling using a low-pass filter ($H_s(f)$) defined in the Fourier space by the $C^{\infty}$ function  (Eq. \ref{eq:Hdef}):

   \begin{equation}
      H_s(f)=\tanh\frac{s\cdot(f + 0.5) + 1}{2}\tanh\frac{s \cdot(-f + 0.5) + 1}{2}.
      \label{eq:Hdef} 
   \end{equation}
   The actual 1D re-sampling kernel $h_k(x)$ in the image domain is obtained computing the inverse Fourier transform of $H_s(f)$  \citep{REBIN}. Since SpectraPy works with 2D images, the 2D re-sampling kernel $w_s$ is obtained combining two separate 1D kernels along the two orthogonal axes: $w_s(x,y) = h_s(x)\cdot h_s(y)$.

\section{Conclusions}
   I have presented the first public release of the SpectraPy package, a Python library to reduce and extract astronomical spectra. SpectraPy wants to be an instrument independent library. This release of the library is focused on the rectification and extraction of 2D spectra. For the future I want to add functionality finalized to remove instrument signatures, like bias subtraction, flat field, pixel to pixel response correction. The idea is to develop an open ecosystem, which collects methods and facilities for spectra extraction and calibration of through-slit astronomical spectra.

\bibliography{C19}

\end{document}